\documentclass[useAMS,usenatbib]{mn2e}
\usepackage{times}
\usepackage{graphicx}
\usepackage{amssymb}

\title[Three supernova shells around a young star cluster in M33]{Three supernova shells around an M33 star cluster}

\author[Camps-Fari\~{n}a et al.]{A. Camps-Fari\~{n}a$^{1,2}$\thanks{E-mail:artemic@iac.es}, John E. Beckman$^{1,2,3}$, J. Font$^{1,2}$ and A. Borlaff$^{1,2}$,\newauthor J. Zaragoza-Cardiel$^{1,2,4}$ and P. Amram$^{5}$\\
$^1$Instituto de Astrof\'{i}sica de Canarias, Tenerife, Spain, E38205\\
$^2$Departamento de Astrof\'{i}sica, Universidad de la Laguna, Tenerife, Spain, E38205\\
$^3$CSIC, Madrid, Spain, 28006\\
$^4$Instituto de Astronom\'{i}a, Universidad Nacional Aut\'{o}noma de M\'{e}xico, D.F. M\'{e}xico\\
$^5$Laboratoire d'Astrophysique de Marseille, Aix Marseille Universit\'{e}, CNRS, F-13388, Marseille, France\\
}

\begin{document}

\pagerange{\pageref{firstpage}--\pageref{lastpage}} \pubyear{2002}
\maketitle
\label{firstpage}

\begin{abstract}
Using a specialized technique sensitive to the presence of expanding ionized gas we have detected a set of three concentric expanding shells in an HII region in the nearby spiral galaxy M33. After mapping the kinematics in H$\alpha$ with Fabry-Perot spectroscopy we used slit spectra to measure the intensities of the [SII] doublet at $\lambda \lambda$ 671.9, 673.1 nm and the [NII] doublet at $\lambda \lambda$ 645.8, 658.3 nm to corroborate the kinematics and apply diagnostic tests using line ratios. These showed that the expanding shells are shock dominated as would be the case if they had originated with supernova explosions. Estimating their kinetic energies we find fairly low values, indicating a fairly advanced stage of evolution. We obtain density, mass and parent star mass estimates, which, along with the kinetic energies, are inconsistent with the simplest models of shock-interstellar medium interaction. We propose that the presence and properties of an inhomogeneous medium offer a scenario which can account for these observations, and discuss the implications. Comparing our results with data from the literature supports the combined presence of an HII region and supernova remnant material at the observed position.
\end{abstract}

\begin{keywords}

HII regions -- ISM: bubbles -- ISM: supernova remnants -- galaxies: star clusters: general

\end{keywords}

\section{Introduction}
It is recognized that feedback from stellar winds and supernovae has had a decisive effect on galaxy formation and evolution \citep{Dekel1986}. Galaxy formation models without strong stellar feedback convert their gas into stars too rapidly by two orders of magnitude (e.g. \cite{Hopkins2011}) compared to observations \citep{Kennicutt1998}, form too many stars by factors up to $10^3$, (e.g. \cite{Faucher2011}) and cannot explain the distribution of heavy elements in the intergalactic medium (e.g. \cite{Wiersma2010}). Approximations to the effects of supernovae on the interstellar medium are now a feature of simulation models (e.g. \cite{Marinacci2014})which, as well as making ad hoc approximations to ensure that the feedback is sufficiently strong to reproduce the properties of observed galaxies, have traditionally assumed as simplifying limitations that the surrounding ISM is homogeneous, and supernovae occur singly \citep{Chevalier1974,Thornton1998}. These limitations have been gradually relaxed, as computing power has increased.

It is known that the real ISM is highly inhomogeneous \citep{Osterbrock1959,McKee1977,Giammanco2004} and this has been used in a recent set of models by \cite{Martizzi2015}. The effect of multiple supernovae occurring within a stellar cluster was modelled by \cite{Kim2011} and more recently by \cite{Hennebelle2014}. \cite{Martizzi2015} produced models with both of these features. The aim of all the models is to reproduce the balance between star formation enhancement and quenching due to the effects of the supernovae on their surroundings, and thereby to predict the macroscopic properties of galaxies. As is frequent in all issues related to star formation, particularly massive star formation, a key difficulty for modellers is the lack of observational constraints; in this article we report observations which take a step forward in improving them.

\section{Observations}
We observed a field on the southern arm of M33 centered on the HII region studied in this article during on 31st October 2014 using the Galaxy H$\alpha$ Fabry-Perot System (GH$\alpha$FaS, see \cite{Hernandez2008,Fathi2008}), and followed this up using the Intermediate dispersion Spectrograph and Imaging System (ISIS) on 25th December 2014.
The region in question is listed as BCLMP17-d, the central part of the BCLMP17 complex, in the 1974 Boulesteix HII region catalogue \citep{Boulesteix1974,Hodge2002}.

GH$\alpha$FaS is an integral field spectrometer mounted at the Nasmyth focus of the 4.2m William Herschel Telescope (WHT) at the Observatorio del Roque de los Muchachos (ORM, La Palma), producing a 3.4x3.4 arcmin data cube with seeing-limited angular resolution (spaxel size $\sim0.2$ arcsec) and a spectrum at each point on the object, with 48 channels covering a spectral range close to $\sim400$ km/s, yielding a nominal velocity resolution of $\sim$8 km/s for H$\alpha$.
The data reduction procedures for GH$\alpha$FaS have been well described in the literature. Here we give a very brief summary. GH$\alpha$FaS does not use an optical derotator so a correction is applied by software, as described in detail in \cite{Blasco2010}. Velocity calibration, phase correction, sky removal and adaptive binning are performed as described in \cite{Daigle2006} producing the final cube used here.

ISIS is a high-efficiency, double-armed, medium-resolution (8-120 \AA{}/mm) spectrograph,at the Cassegrain focus of the WHT, capable of providing up to ~4' slit length and ~22" slit width. We observed the region with a 10'' wide slit using the red arm R1200R grating centred at 6800 \AA. This allowed us to observe simultaneously the H$\alpha$, SII and NII lines with a spectral resolution of $\sim$12 km/s. We used a slit width of 10'' to encompass in a single observation the central part of the region where we detected the bubbles with GH$\alpha$FaS.

We reduced the spectrum using \textit{IRAF}. As we use only line flux ratios no flux calibration was necessary.

\section{Expansion maps}
To search for expansive components we analysed the GH$\alpha$FaS data cube with \textsc{bubbly} \citep{Camps2015}, a program designed to find and fit multi-component spectral line data cubes. The analysis is automatic, using a mathematical property of Gaussians to detect multiple components, which are then fitted (if they are 
above a threshold signal-to-noise) and checked to look for pairs of components equidistant in velocity (within uncertainty limits) from the central peak representing the bulk of the HII region. These detections are then mapped in expansion velocity giving "expansion maps", which we use to find the bubbles.

\begin{figure}
\includegraphics[width=\linewidth]{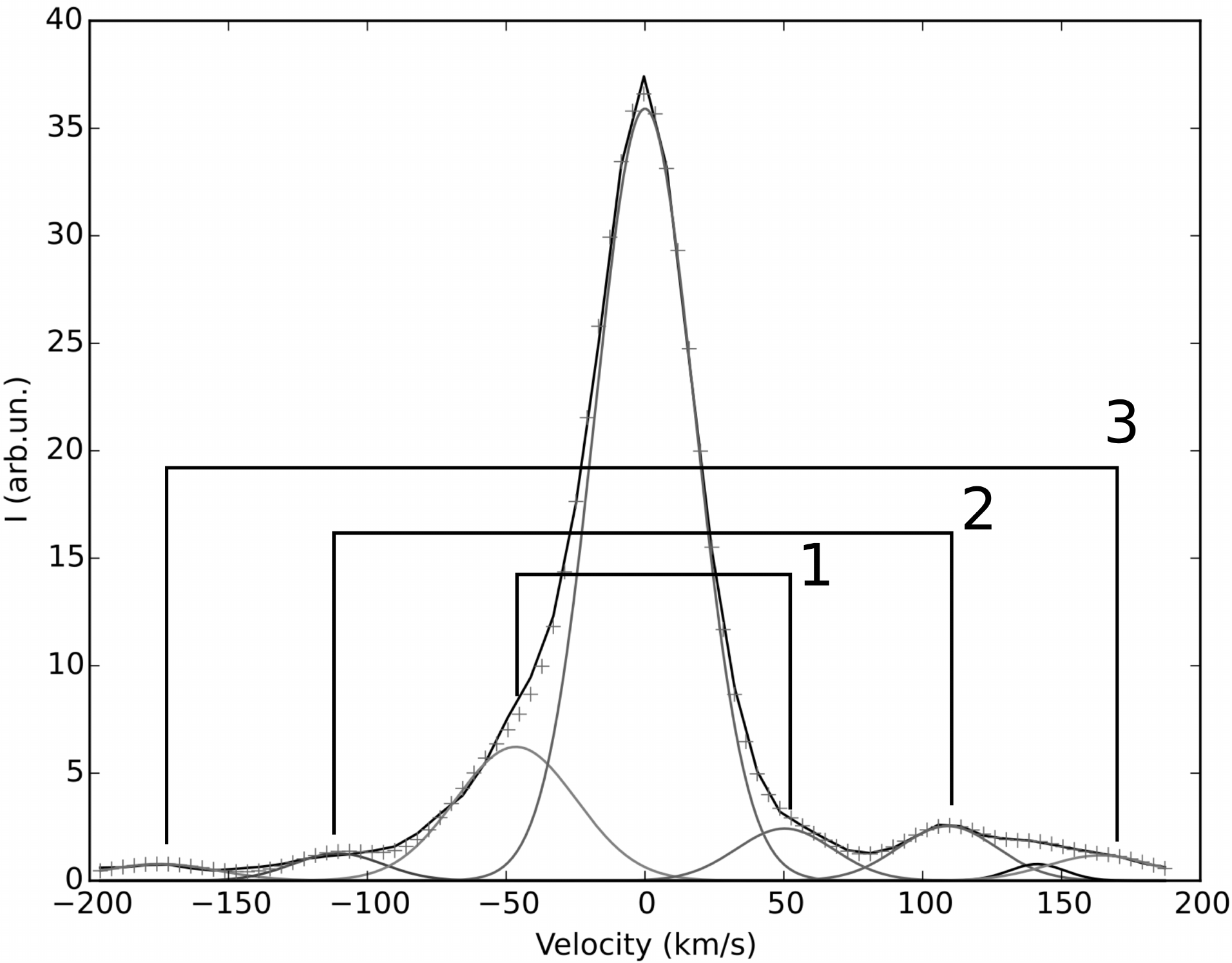}
\caption{Example of a spaxel from the GH$\alpha$FaS data cube which shows three pairs of symmetrically placed secondary peaks, indicating expansion. The detection and fitting was performed automatically using \textsc{bubbly}.}
\label{fig:prof}
\end{figure}

We used the default configuration in the \textsc{BUBBLY} parameter file with a minimum S:N ratio of 5 to detect the expansion components. Because of the relatively low signal-to-noise ratio threshold we added visual inspection of the expansion maps to ensure that detection was spatially coherent.
In Fig. \ref{fig:prof} we show the line profile of H$\alpha$ emission which indicates the presence of the three shells, with a central main peak and three pairs of secondary components. This profile comes from only a single resolved element of the velocity map.

\begin{figure*}
\includegraphics[width=0.85\linewidth]{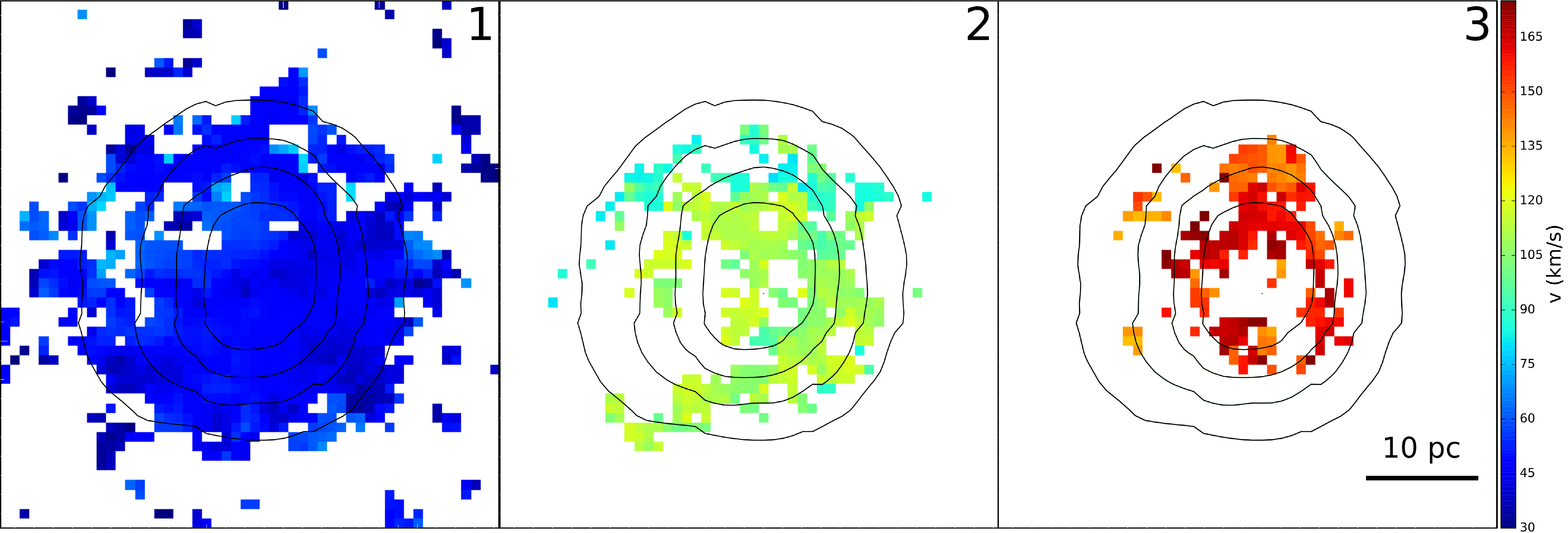}
\caption{Expansion maps of the three detected bubbles, which show the detected expansion velocity in each pixel, all in the same velocity scale. Overlaid are contours of the region's H$\alpha$ emission, it can be seen that the bubbles are roughly concentric with each other and the region}
\label{fig:expmaps}
\end{figure*}

Figure \ref{fig:expmaps} was produced by mapping all the pixels where a given pair of peaks was detected, which displays the projected extent of the shell; the process was repeated for each of the three shells. We can see that the smallest shell has the most rapid expansion velocity, and the largest shell has the slowest, as expected for shells originating in the centre and doing work on the surrounding interstellar medium during their expansion. 

\section{Properties of the bubbles}
We can measure the radius and expansion velocity directly on the expansion maps and, using the H$\alpha$ flux of the shells, can calculate their density\citep{Relano2005}:
\begin{equation}
L_{H{\alpha}}(shell)=4\pi R^2 \Delta R \, n_{e}^2 \alpha(H,T) h \nu
\label{lum_eq}
\end{equation}

We cannot independently calculate the shell width $ \Delta R$, so we took a lower limit arising from the evolution of a strong shock disregarding instabilities\citep{Lozinskaya1992}. A lower limit in shell thickness means a lower limit in the total mass of the shell. The fact that we measure a lower limit in mass will be important later in the discussion.

The flux is measured calculating the fraction of flux the pair of secondary components emits and using a calibrated H$\alpha$ image of M33 \citep{Hoopes2001}. We chose this method because we are more confident in the calibration of the photometry of this image rather than calibrating GH$\alpha$FaS.

We can also calculate mass and energy and using these three properties:
\begin{eqnarray}
{M=4\pi R^2 \Delta R \: n_e m_p} &{;}& {E_k=\frac{1}{2}M*v_{exp}^2}
\end{eqnarray}

The age is estimated using the relation of $t=2/7*(R/v)$, from the analytic expression for the evolution of a supernova remnant in the snowplough phase \citep{McKee1977}. Using this expression is justified in the following sections, where we also discuss the validity of the values calculated.

Lastly, we calculate the mean pre-shock density for each bubble assuming the shells swept out all of the gas they encountered, dividing the total mass of the shell by the total spherical bubble volume. We disregard the masses of the SN ejecta (a few solar masses), as the shell masses have much larger values.

The ISIS spectrum was used to compute emission line ratios, to probe the origin of the shells. To obtain results for each shell separately we applied a modified version of \textsc{bubbly} to each spectrum over the region in the spatial direction for each of these lines: H$\alpha$, [SII]$_{6716}$, [SII]$_{6731}$ and [NII]$_{6584}$. To compute the doublet [NII] we used the relation $[\mathrm{NII}]_{6584} = 3 *[\mathrm{NII}]_{6548}$ which is independent of the physical parameters of the gas.

To improve the statistical significance of the ratios we repeated the detection using binning of 2, 3, 5, 6, 15 and 30 elements in the spatial direction, and used the mean value with the standard deviation as the error after eliminating clearly discrepant values.

We could in principle also use the [SII]$_{6716}$/[SII]$_{6731}$ ratio as an indicator of density, but unfortunately the values measured are not within the range in which the ratio is sufficiently sensitive to density. 
Table \ref{table_prop} lists all parameters measured in the bubbles, both from GH$\alpha$FaS and ISIS data.

\begin{table}
\centering
\caption{Physical properties of the bubbles}
\begin{tabular}{lcccc}
\hline \hline
& Bubble 1 & Bubble 2 & Bubble 3\\
Radius    & 22 $\pm$ 2 & 16 $\pm$ 2 & 12.5 $\pm$ 1 \\
(pc) &  &  &  \\
$v_{exp}$ & 55 $\pm$ 5 & 115 $\pm$ 12 & 165 $\pm$ 17 \\ 
(km/s) &  &  &  \\ 
$L_{H\alpha}$  & 4.4 $\pm$ 0.4 & 1.4 $\pm$ 0.2 & 0.61 $\pm$ 0.06 \\
($10^{36}$ erg/s) & & & \\
n$_{e}$  & 10.2 $\pm$ 1.1 & 9.3 $\pm$ 1.1 & 8.8 $\pm$ 1.0\\
(cm$^{-3}$) & & & \\
Mass  & 337 $\pm$ 34& 118 $\pm$ 12& 54 $\pm$ 5 \\
(M$_\odot$) & & & \\
E$_k$   &  1.0 $\pm$ 0.2 & 1.6 $\pm$ 0.3 & 1.4 $\pm$ 0.3\\
($10^{49}$ erg)& & & \\
Age*  & 114 $\pm$ 16 & 40 $\pm$ 6 & 21 $\pm$ 3 \\
(kyr)& & & \\
n$_0$* & 0.31 $\pm$ 0.06 & 0.28 $\pm$ 0.06 & 0.26 $\pm$ 0.05\\
(cm$^{-3}$)& & & \\
$[\mathrm{SII}]/\mathrm{H}\alpha$ & 0.4 $\pm$ 0.15 & 0.45 $\pm$ 0.16 & 0.43 $\pm$ 0.09 \\
$[\mathrm{NII}]/\mathrm{H}\alpha$ & 0.44 $\pm$ 0.2 & 0.35 $\pm$ 0.09 & 0.27 $\pm$ 0.13 \\
$[\mathrm{SII}]_1/[\mathrm{SII}]_2$ & 1.38 $\pm$ 0.25 & 1.41 $\pm$ 0.18 & 1.42 $\pm$ 0.26 \\
\hline
\end{tabular}
\\
\begin{flushleft}
\textbf{Notes.} See section 4 for details on how the properties were determined. n$_{e}$ is the electron density inside the shell, while n$_0$ is the average ambient particle density it encountered. The properties marked with * (Age and ambient density) are only valid for a shock evolving in a homogeneous medium. The calculations use a lower limit for shell thickness, this implies that mass and kinetic energy are lower limits while density is an upper limit. Even then, the dependence on shell thickness is $\sqrt{\Delta R}$, so a factor 10 increase in the thickness would increase mass and energy a factor $\sim$ 3.
\end{flushleft}
\label{table_prop}
\end{table}

\subsection{Archival properties of the cluster}
We have supporting information from the literature obtained by querying the position with VizieR; the presence of a supernova remnant in this HII region detected using radio observations at lower angular resolution is reported by Gordon et al. \citep{Gordon1999} (Tables 1,2 \#90). Also relevant is a survey of HII regions and molecular gas by \cite{Miura2012} which used archival Hubble Legacy Archive images to date the accompanying star cluster by fitting isochrones to a color-magnitude diagram extracted using aperture photometry (see section 3.2.2). This gives an age range of 6-12 Myr for the cluster, which is also reported to have 51 O stars (Table 3 \#27). This survey also shows a $2.2\cdot 10^6$ M$_\odot$ molecular cloud adjacent to the cluster (Table 2 \#24).
\subsection{Identification as supernova remnants}
\begin{figure}
\includegraphics[width=\linewidth]{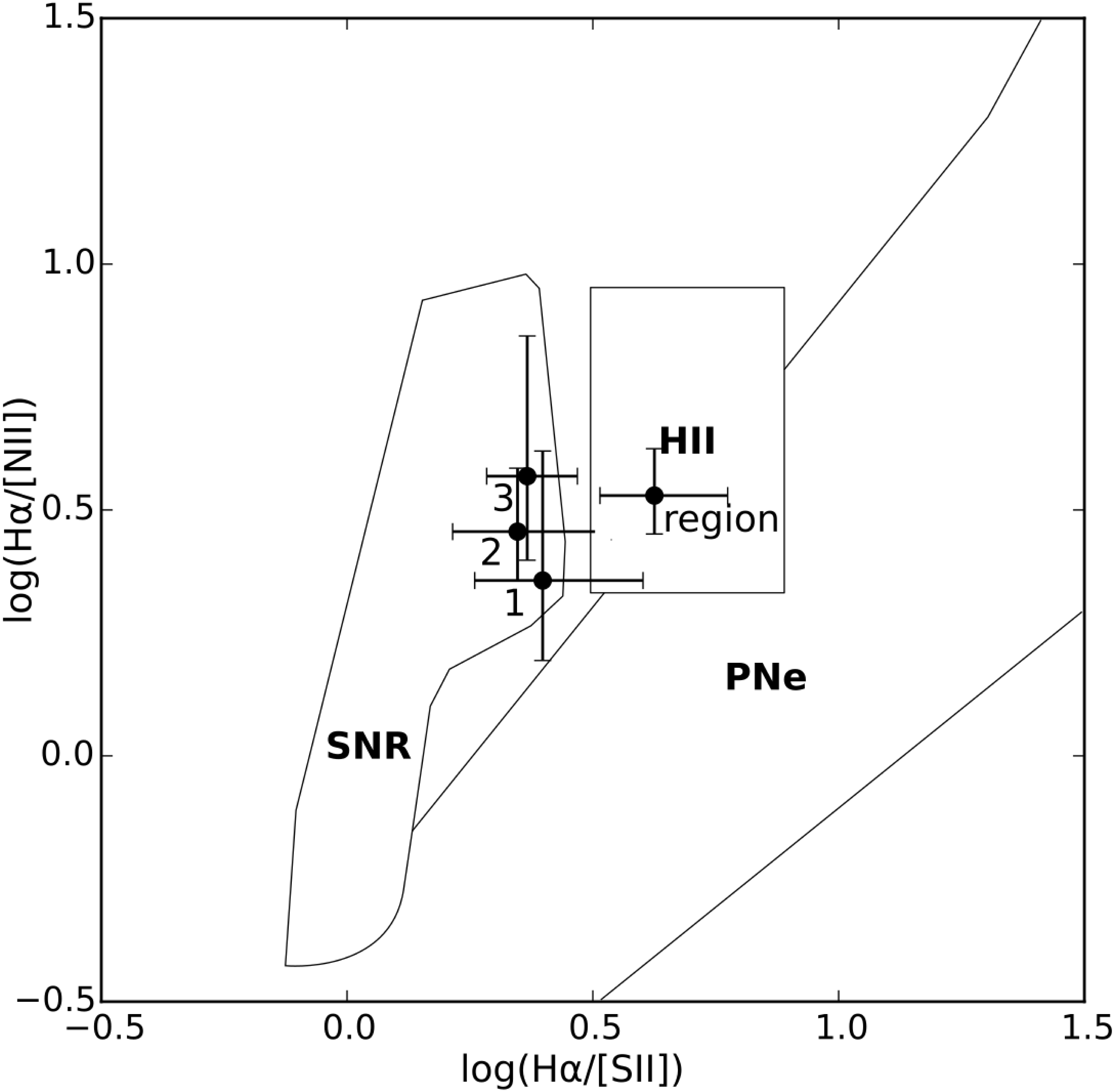}
\caption{Diagnostic diagram used to distinguish between supernova remnants (SNR), HII regions (HII) and planetary nebulae (PNe) based on the relation between the SII to H$\alpha$ and NII to H$\alpha$ line ratios. We plot the remnants with their identifier numbers as well as the HII region.}
\label{fig:lineratios}
\end{figure}

The identification as supernova remnants was made using forbidden line ratios [SII]/H$\alpha$ and [NII]/H$\alpha$ to place our bubbles within diagnostic plots which distinguish between supernova remnants, HII regions and planetary nebulae \citep{Sabbadin1977,Garcia1991}. The classic diagnostic of [SII]/H$\alpha$ $>$ 0.4 is included in these plots as the region for SNR is delimited by this value on the X axis (see Fig. \ref{fig:lineratios}). All three fall well within the defined regions for SNR, though for bubbles 1 and 2 the error reaches the area for photoionized gas. Significantly, the same analysis for the central peak places it, as expected, within the area for HII regions.

Though the ratios indicate otherwise, an alternative origin for the bubbles might be stellar winds, but this is ruled out in practice; as winds are continuously injecting energy since the stars' birth they cannot produce the segregated concentric structure observed, as the individual bubbles would blend and merge into one structure before growing to their current size, which already encompasses most of the cluster. The only possibility for winds  to produce bubbles afterwards would be a red supergiant wind accelerated by a subsequent Wolf-Rayet phase (see \cite{Freyer2006}), but the stars capable of producing this effect have masses over 25 M$_{\odot}$ and could no longer be present, barring multiple star formation bursts.

\subsection{Physical properties}

The evolution of supernova remnants is divided into three main phases, an initial free expansion while the mass of the ejecta dominates, an adiabatic expansion once the swept-up material mass is similar to that of the ejecta and a radiative phase, also called snowplough phase, in which the losses by radiation become significant and the gas cools. Given that the initial energy of a supernova is around $10^{51}$ erg our remnants currently have lost some 99\% of their energy, which means that they must have been radiative for some time. We can check this using the expressions for the adiabatic phase \citep{Hnatyk2007}:

\begin{eqnarray}
{R_{ST} (t) = \left(  \frac{25 E_{SN}}{4 \pi \rho_0}   \right)^{1/5}} &{;}& {t_{tr}=2.84 \cdot 10^4 E_{51}^{4/17} n_0 ^{-9/17}}
\end{eqnarray}

where the left expression is the evolution of the radius and on the right we have the transition time to radiative phase. For our measured ambient density we find a transition time of 54 kyr which already does not fit the data, two of our remnants should not have transitioned yet. Furthermore, even with the faster expansion of the adiabatic phase we find predicted sizes roughly double the measured radii. We will address this apparent inconsistency again shortly.

The mass range for the progenitor stars, calculated from the age of the cluster, is from $\sim$15 to 19 M$_{\odot}$, which is not unreasonable if we find three similar mass stars in a cluster of this size. Given the relatively short intervals between the supernova explosions, their progenitors must have had very similar masses.

The swept-up masses of the shells are very interesting. Their values are not consistent with the simplest models but they give us a clue to solve the inconsistency in the physical parameters. In a homogeneous medium the expanding shock should sweep out almost all the surrounding gas into the shell leaving a very rarefied medium, so that a second supernova soon after the first should sweep up hardly any material, lengthening the free-expansion phase. In spite of this, we find that each consecutive remnant has accumulated tens of solar masses of material. This is so even though our value for the thickness of each shell is a lower limit, meaning the masses are also lower limits as they scale with the square root of the thickness. A mechanism is needed to replenish gas inside a remnant for the next shell to encounter sufficient material; the most plausible is mass-loading from cold dense clouds. It has long been suspected that the interior of an HII region is not homogeneous nor fully ionized, and that small molecular clouds can survive for a long time inside it as they slowly evaporate \citep{Bok1947}.

Hydrodynamic ablation by shocks or supersonic winds can also disrupt such clouds, so each supernova shell feeds gas to the next one.
The effects of the interaction on the shell are also important. Simulations show \citep{Chevalier1974} that upon encountering a cold, dense cloud, the shell will tend towards radiative evolution, rapidly slowing down and losing energy, greatly advancing its stage of evolution. In the cited paper a simple model with a single density jump at a fixed radius is used, while we expect clusters of cloudlets instead. Either way, the effect of a density jump on the shell is what we have observed in our remnants, so adding the condition of inhomogeneity to our model explains the apparent inconsistencies. It also means that the ages estimated assuming a homogeneous medium are not correct. An inhomogeneous medium slows down the remnants, so our values are upper limits, but they should be within a small factor of the real expansion ages.

\section{Discussion}
We have presented results which seem to indicate the presence of three currently expanding supernova remnants within a single young HII region. Our key claim is that the expansion detected must arise from three different objects. Initially, we considered other geometrical configurations, such as typical complex structures in HII regions with several arches and cavities, but we detect very clearly an approaching and receding component for all three along the same lines of sight, so they are either nested or aligned on the line of sight. Given the similar size of the bubbles relative to the cluster, and their concentricity, we believe that nested shells offer the simpler explanation.

We found that the physical properties appeared inconsistent given their masses, kinetic energies and ages. This is solved considering an inhomogeneous medium with embedded dense clumps of molecular gas, which would rob the remnants of energy as well as refill the ambient gas to be swept up by the subsequent supernova explosions. Because of this, our interpretation requires such clumps to survive for a long time inside the HII region.

The survival of clumps inside HII regions has long since been documented, \citep{Bok1947}, and they have been found even in extreme examples such as 30 Doradus \citep{Indebetouw2013}, where a complex of molecular clouds is found associated with the brightest star cluster.

The effect of stellar winds and radiation on their parent molecular clouds has been studied in simulations by \cite{Rogers2013} and \cite{Dale2014}. In the former, three massive $\sim$30 M$_{\odot}$ stars affect a surrounding 4 pc massive molecular cloud. The main finding is that the winds carve low density channels through which the gas escapes, making the denser regions surprisingly resistant to ablation. Another important result is that dense clumps are largely impervious to supernova shocks. They find that the dense material is either destroyed or pushed outside the original 4 pc radius in just over 6 Myr, which coincides with the lower limit of the age for our HII region. 
\cite{Dale2014} perform cluster-scale simulations and find results compatible to those of \cite{Rogers2013} with a very structured dense medium. They use molecular clouds with a variety of physical parameters finding that the denser, more massive clouds are much harder to disrupt. For lower mass clouds the stars carve irregular cavities on scales of 10 pc. The molecular cloud associated with our cluster is similar to the densest and most massive of \cite{Dale2014}'s.

Both studies show that the presence of embedded clumps is expected in clusters born from massive molecular clouds, and that they can easily survive until the stars explode as supernovae.
We also need to estimate the rate of cloud evaporation required to refill the gas and reproduce the masses in the observed shell. From the masses and ages of the remnants we require an injection rate of about 25 M$_{\odot}$ per 10 kyr. \cite{Dale2014} find that except at the onset of star formation, winds hardly affect the disruption of the molecular cloud; with photoevaporation dominating this process, so here we need consider only cloud destruction by photoevaporation.

Using the relation from \cite{Pittard2007} for photoevaporation and taking a sample cloud with radius 1 pc at a distance of 10 pc from the centre of the cluster, we estimate the flux of ionizing photons from the cluster using its H$\alpha$ luminosity and obtain an injection rate of 0.8 M$_{\odot}$ per 10 kyr. We can therefore easily reproduce the required mass injection without an overly large number of clouds, even accounting for the fact that our masses are lower limits.

Feedback from stellar winds and supernovae plays an important role in many processes relevant to galaxy evolution. As well as those mentioned above there is the possible dissipation of nuclear dark matter cusps \citep{Pontzen2012}, the dissipation of metals both within galaxies \citep{Spitoni2009} and from galaxies into the intergalactic medium \citep{Heckman1990}, and the enhancement of the infall rate of low metallicity intergalactic gas to the disc, by its interaction with supernova ejectae in the galactic halo \citep{Marasco2012}. To incorporate any of these effects into models of the formation and evolution of galaxies, it is clearly important to have not only model inputs to the dynamics of the interaction of the winds and supernovae with the surrounding ISM, but measurements which can help quantify the process. As the stars in massive star clusters form virtually simultaneously, cases of multiple supernova explosions must be common enough to be incorporated into the relevant models.But much more generally to make realistic models mass loading of the expansion via the dissipation of dense cool clumps must be taken into consideration. The triple supernova has allowed us to illustrate its importance, and make quantitative estimates of relevant parameters for much wider application.

\section*{Acknowledgements}
We thank M. Beasley, J.E. Dale, and the anonymous referee for useful discussion and comments.
This research has been supported by the Instituto de Astrof\'{i}sica de Canarias under project P/308603. JEB acknowledges financial support to the DAGAL network from the People Programme (Marie Curie Actions) of the European Union's Seventh Framework Programme FP7/2007-2013/ under REA grant agreement number PITN-GA-2011-289313.

Based on observations made with the William Herschel Telescope under Director's Discretionary Time of Spain's Instituto de Astrof\'{i}sica de Canarias.

The William Herschel Telescope is operated on the island of La Palma by the Isaac Newton Group in the Spanish Observatorio del Roque de los Muchachos of the Instituto de Astrof\'{i}sica de Canarias.

This research has made use of the VizieR catalogue access tool, CDS, Strasbourg, France. The original description of the VizieR service was published in A\&{}AS 143, 23.
%This research has made use of NASA's Astrophysics Data System.

\label{lastpage}

\end{document}